# Photochemical and RadiatiOn Transport model for Extensive USe (PROTEUS)

Yuki Nakamura[1,2], Naoki Terada[1], Shungo Koyama[1], Tatsuya Yoshida[1], Hiroki Karyu[1], Kaori Terada[1], Takeshi Kuroda[1], Arihiro Kamada[1], Isao Murata[1], Shotaro Sakai[1], Yuhei Suzuki[1], Mirai Kobayashi[1], and François Leblanc[2]

[1] Graduate School of Science, Tohoku University, Sendai, Japan
[2] LATMOS, Sorbonne Université, Paris, France

Corresponding author: Yuki Nakamura (yuki.nakamura.r2@dc.tohoku.ac.jp)

**Abstract**

We introduce a new flexible one-dimensional photochemical model named Photochemical and RadiatiOn Transport model for Extensive USe (PROTEUS), which consists of a Python graphical user interface (GUI) program and Fortran 90 modules. PROTEUS is designed for adaptability to many planetary atmospheres, for flexibility to deal with thousands of or more chemical reactions with high efficiency, and for intuitive operation with GUI. Chemical reactions can be easily implemented into the Python GUI program in a simple string format, and users can intuitively select a planet and chemical reactions on GUI. Chemical reactions selected on GUI are automatically analyzed by string parsing functions in the Python GUI program, then applied to the Fortran 90 modules to simulate with the selected chemical reactions on a selected planet. PROTEUS can significantly save the time for those who need to develop a new photochemical model; users just need to write chemical reactions in the Python GUI program and just select them on GUI to run a new photochemical model.

**Introduction**

Photochemical models are essential for investigating the vertical chemical structure of planetary atmospheres and their evolution throughout the history of the planets. They solve continuity-transport equations considering production and loss of each atmospheric species by numerous chemical reactions including photolysis. So far, plenty of photochemical models have been developed for various planetary atmospheres (e.g., Kasting et al., 1979; Nair et al., 1994; Kim and Fox, 1994; Fox and Sung, 2001; Krasnopolsky, 2009; Krasnopolsky, 2012; Chaffin et al., 2017). As the mass and spectral resolutions of measurements for detecting chemical species in planetary atmospheres increase and the theory of chemical kinetic systems become more complex, the need for photochemical models with hundreds or thousands of chemical reactions increase.

There are roughly three approaches to develop a numerical code for solving a lot of chemical reactions (Damian et al., 2002). The first approach is a hard-coding approach, in which the developer analyzes the chemical reactions, derives all the production and loss rate terms for each chemical species, and codes them into a program by hand. This approach is easy to develop when the number of chemical reactions is smaller than a hundred, however, it takes a lot of time to develop a code when the number of reactions becomes larger than hundreds or more. It is also difficult to add new chemical reactions into an already hard-coded program.

The second approach is a totally integrated approach, in which the chemical reactions are listed in a specific file in a certain format and they are parsed by a program and stored in a memory when it is run. This approach is flexible in adding new chemical reactions after the development of the core program, and easy to deal with hundreds of chemical reactions without developer's manual derivation. This approach was used in the Atmos model in Fortran language for instance, which was originally developed by Kasting et al. (1979), updated by Zahnle et al. (2006) and recently described in Arney et al. (2016). Recently an integrated Martian photochemical model was developed by Chaffin et al. (2017) in Julia language with a more flexible way of describing reactions and rate coefficients.

The third approach is a preprocessing approach, in which the chemical reactions are listed in a specific file like the totally integrated approach, but are parsed by a preprocessor to generate a hard-coded program in a high-level language such as Fortran or C language (Damian et al., 2002). This approach is also flexible in implementing hundreds of chemical reactions and is as efficient as the hard-coding approach. This approach was used in the kinetic preprocessor (KPP) originally developed by Damian et al. (2002), which has been widely used for chemical kinetic models for Earth's atmosphere.

In this paper, we present a new integrated photochemical model named Photochemical and RadiatiOn Transport model for Extensive USe (PROTEUS), with a totally integrated approach. PROTEUS couples Python and Fortran modules, which is designed for adaptability to many planetary atmospheres, for flexibility to deal with thousands of or more chemical reactions with high efficiency, and for intuitive operation with a graphical user interface (GUI). A Python GUI program integrates a list of chemical reactions, GUI functions controlling the behavior of GUI operation, and a string parsing functions analyzing chemical reactions that output a Fortran 90 module. Fortran 90 modules solve differential equations numerically. Chemical reactions are written in a simple and flexible string format in the Python GUI program, making it easy to add new chemical reactions into the Python GUI program. The feature of PROTEUS that the Python GUI program outputs a Fortran 90 module is similar to the preprocessing approach, leading to a high efficiency, however, the Fortran modules in PROTEUS are not hard-coded but is rather generic.

PROTEUS has been newly developed and independent of other photochemical models or KPPs that have been developed so far.

**Model description**
**Equations**
PROTEUS is a one-dimensional photochemical model that solves a system of continuity equations for each species as follows:

$$\frac{\partial n_i}{\partial t} = P_i - L_i - \frac{\partial \Phi_i}{\partial z}, \quad (1)$$

where $n_i$ is the number density of $i$th species, $P_i$ is the production rate of $i$th species, $L_i$ is the loss rate of $i$th species, $z$ is the altitude and $\Phi_i$ is the vertical flux of $i$th species. The vertical flux $\Phi_i$ for both neutral and ionized species can be expressed as follows:

$$\Phi_i = -n_i D_i \left( \frac{1}{n_i} \frac{\partial n_i}{\partial z} + \frac{1}{H_i} + \frac{q_i}{q_e} \frac{T_e/T_i}{P_e} \frac{\partial P_e}{\partial z} + \frac{1+\alpha_i}{T_i} \frac{\partial T_i}{\partial z} \right) - n_i K \left( \frac{1}{n_i} \frac{dn_i}{dz} + \frac{1}{H} + \frac{1}{T} \frac{dT}{dz} \right), \quad (2)$$

where $D_i$ is the binary diffusion coefficient between $i$th species and the background atmosphere, $H_i = k_B T_i / m_i g$ is the scale height of $i$th species, $m_i$ is the mass of $i$th species, $k_B$ is the Boltzmann constant, $g$ is the gravitational acceleration, $q_i$ is the charge of $i$th species, $q_e$ is the elementary charge, $T_e$ and $T_i$ are the temperatures of electrons and $i$th species, respectively, $P_e = n_e k_B T_e$ is the electron pressure, $n_e$ is the electron number density, $\alpha_i$ is the thermal diffusion coefficient, $K$ is the eddy diffusion coefficient, $H = k_B T / m g$ is the mean scale height of the background atmosphere, $m$ is the mean molecular mass of the atmosphere, and $T$ is the neutral temperature. The temperature profiles are assumed to be stationary in time. The third term in Equation (2) is the ambipolar diffusion term, which is applied only to charged species. The altitude-dependent gravitational acceleration $g$ is calculated by using the mass and radius of the planet. The basic equations are stiff equations in which some of the variables such as number densities of short-lived species change more quickly than others. Thus, PROTEUS applied an implicit method solving the differential equations as follows:

$$\boldsymbol{x}^{n+1} = \boldsymbol{x}^n + \left(\frac{\mathbf{I}}{\Delta t} - \boldsymbol{J}\right)^{-1} \boldsymbol{F}(\boldsymbol{x}^n) \quad (3)$$

where $\boldsymbol{x}^n$ and $\boldsymbol{x}^{n+1}$ are vector forms of the number density of chemical species at time step $n$ and $n+1$, respectively, $\mathbf{I}$ is the unit matrix, $\Delta t$ is the timestep size, $\boldsymbol{F}(\boldsymbol{x}^n)$ is the right-hand side of equation (1) in the vector form, and $\boldsymbol{J}$ is the sparse Jacobian matrix defined as follows:

$$\boldsymbol{J} \equiv \frac{\partial \boldsymbol{F}(\boldsymbol{x}^n)}{\partial \boldsymbol{x}^n} = \begin{pmatrix} \frac{\partial F_1}{\partial x_1} & \frac{\partial F_1}{\partial x_2} & \cdots & \frac{\partial F_1}{\partial x_N} \\ \frac{\partial F_2}{\partial x_1} & \frac{\partial F_2}{\partial x_2} & \cdots & \frac{\partial F_2}{\partial x_N} \\ \vdots & \vdots & \ddots & \vdots \\ \frac{\partial F_N}{\partial x_1} & \frac{\partial F_N}{\partial x_2} & \cdots & \frac{\partial F_N}{\partial x_N} \end{pmatrix} \quad (4)$$

The time step size can gradually increase from an initial value $10^{-8}$ sec to ideally more than $10^{14}$ sec, allowing us to investigate from a sporadic event response in several minutes to the evolution of planetary atmospheres in a billion-year time scale.

PROTEUS is basically a one-dimensional photochemical model and currently does not solve horizontal transportation, but considers the rotation of a planet as options to obtain a simplified global distribution. PROTEUS has four options for the simulation geometry: (1) one-dimensional simulation at a given latitude at noon, (2) two-dimensional simulation at noon at each latitude from the north pole to the south pole, (3) two-dimensional simulation at a given latitude with rotation, and (4) three-dimensional simulation at all latitudes with rotation. The three-dimensional simulation has already been applied by Nakamura et al. (2022) for the Jovian ionosphere.

**Radiative transfer**
PROTEUS uses the solar EUV irradiance model for aeronomic calculations (EUVAC) (Richards et al., 1994) for the reference irradiance spectrum of solar extreme ultraviolet (EUV) flux to calculate the photoionization rates of atmospheric species. EUVAC model provides the solar EUV flux in 37 wavelength bins ranging from 5 nm to 105 nm. EUVAC model requires the input of F10.7 value and its 81 days running average value. High resolution solar EUV reference flux model such as the high-resolution version of EUVAC (HEUVAC) (Richards et al., 2006) and the Flare Irradiance Spectral Model-Version 2 (FISM2) (Chamberlin et al., 2020) will be implemented into PROTEUS in the future. For calculating the photodissociation rates of atmospheric species, we used the reference irradiance spectrum of the solar flux in the wavelength range 0.05-2499.5 nm taken from Woods et al. (2009). Adopting the solar flux taken from EUVAC model and Woods et al. (2009), the radiative transfer is solved by considering the absorption of the solar irradiation by atmospheric species. In PROTEUS, users can flexibly change the wavelength bin size for the solar irradiance of Woods et al. (2009) and absorption/dissociation cross sections of chemical species at each wavelength. The solar flux and cross section data can be provided in any wavelength bins, which are automatically interpolated and binned to the wavelength bin given by the user. The automatically binning algorithm is especially useful when users need high resolution wavelength bins in a limited wavelength range. For example, if users need to resolve the Schumann-Runge bands of the oxygen molecule, the user can set a 0.01 nm resolution at 176-192.6 nm and 1 nm at other wavelength range, which could reduce the computational cost in solving radiation transfer and dissociation rate of atmospheric species even fully resolving the structured Schumann-Runge bands of the oxygen molecule. This algorithm is also useful for resolving a slight difference in absorption cross sections between isotopes (Yoshida, et al., 2022 submitted).

The reference solar irradiance spectra are then divided by the square of the distance $r$ in the unit AU between the planet and the Sun. The distance $r$ between the planet and the Sun at a given solar longitude $L_s$ is given by

$$r = r_m \frac{1 - e^2}{1 + e \cos(L_s - L_{s,P})} \quad (5)$$

where $r_m$ is the mean distance in unit AU between the planet and the Sun, $e$ is the eccentricity of the planetary orbit, and $L_{s,P}$ is the solar longitude at perihelion. The solar zenith angle at latitude $\theta$ and an hour angle $\eta$ is given by

$$\cos \chi = \sin \theta \sin \delta + \cos \theta \cos \delta \cos \eta \quad (6)$$

where $\delta$ is solar declination. $\delta$ and $\eta$ are given by

$$\sin \delta = \sin \varepsilon \sin L_s \quad (7)$$

$$\eta = \frac{2\pi t_L}{T_p} \quad (8)$$

where $\varepsilon$ is the tilt angle of the rotational axis, $t_L$ is the time in second measured from the local noon, and $T_p$ is the rotational period of the planet.

The absorption, ionization, and dissociation cross sections implemented into PROTEUS are described in the following sections and Appendix. The Rayleigh scattering cross section $\sigma_R$ is given by (Liou, 2002):

$$\sigma_R = \frac{128\pi^5}{3\lambda^4} \alpha_p^2 \quad (9)$$

where $\lambda$ and $\alpha_p$ are the wavelength in nm and polarizability in nm$^3$ of gaseous species, respectively. The polarizability $\alpha_p$ of gaseous species are taken from the Computational Chemistry Comparison and Benchmark DataBase (https://cccbdb.nist.gov).

**Structure of PROTEUS**

PROTEUS consists of a Python GUI program and of Fortran 90 modules. The Python GUI program contains a list of chemical reactions for each planet, string parsing functions that parse the reactions and reaction rate coefficients selected in GUI, and GUI functions that control the behavior of GUI. Python language was adopted because of its flexibility in parsing strings and its capability in operating GUI. Since Python language is not efficient for numerical calculation, Fortran language was adopted to solve differential equations numerically. The structure of PROTEUS is illustrated in Figure 1. Chemical reactions listed in the Python file are first read by GUI functions (arrow 1 in Figure 1). Then, users select reactions on GUI (arrows 2 in Figure 1), which will be analyzed by string parsing functions (arrow 3 in Figure 1) to output a Fortran 90 module named "v__in.f90" and data files (in directories "PLJ_list" and "settings") including information of the selected chemical reactions (arrow 4 in Figure 1). The Fortran 90 module "v__in.f90" is the only module that includes information of chemical reactions, and other Fortran 90 modules are independent of chemical reactions to be used in the simulation. Datafiles of initial density profiles and temperature profiles are read by "v__in.f90" (arrow 5 in Figure 1). The selected reactions, settings such as temperature profile and initial density profiles are applied to the Fortran 90 model when the main Fortran 90 routine "e__main.f90" call subroutines in "v__in.f90" (arrow 6 in Figure 1). Users can then run the Fortran model by compiling all the Fortran 90 modules (indicated as "7" in Figure 1).

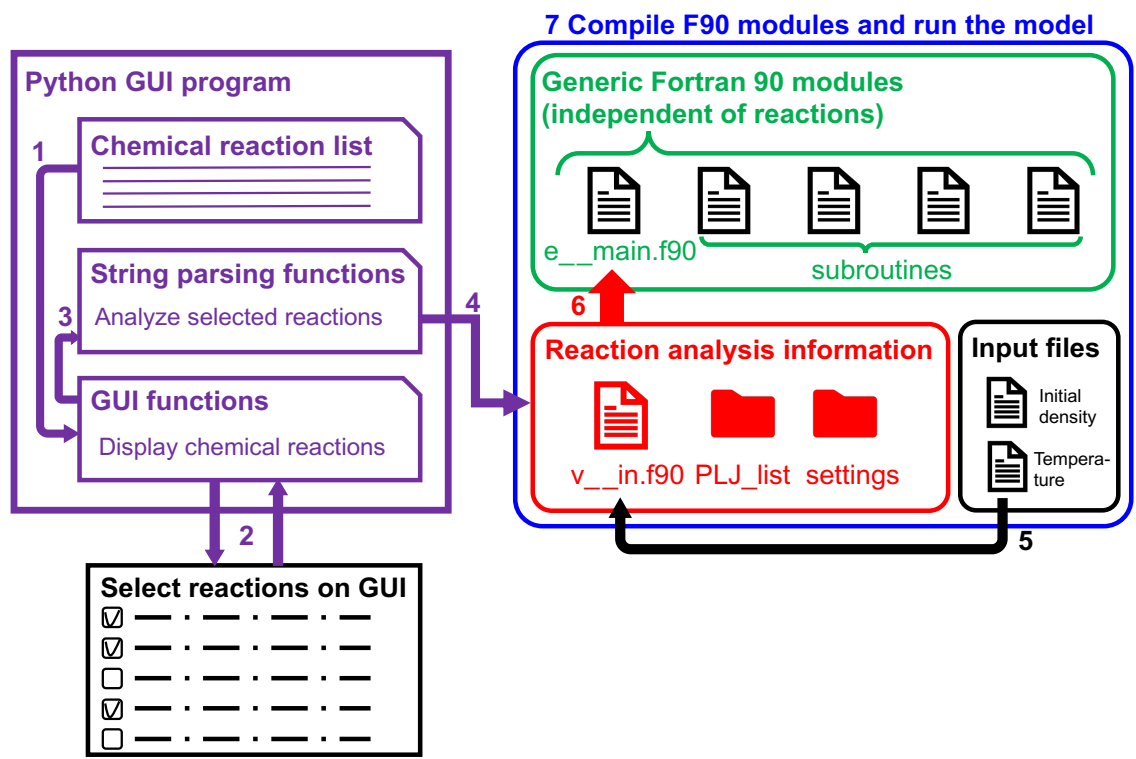

Figure 1 Schematic illustration of the structure of PROTEUS.

The structure of the Fortran codes is illustrated in Figure 2. It should be noted that each Fortran 90 file contains several modules with distinct functions, but only the Fortran 90 file is indicated for simplicity. The main routine "e__main.f90" consists of three parts, (1) initialization, (2) calculation, and (3) finalization. The description of each parts and each Fortran 90 files are as follows.

(1) All variables are defined in "v__tdec.f90". Information of the chemical reactions, boundary conditions, and calculation settings are defined in "v__in.f90". Physical constants and parameters of the planetary orbit are given in "c__prm.f90". Production rates calculated by other models (e.g., ionization rate calculated by a meteoroid model (Nakamura et al., 2022)) can be input in "p__Mars.f90" and "p__Jupiter.f90". The solar EUV flux is calculated by the EUVAC model and the absorption and ionization cross sections are defined in "p__EUVAC.f90". The solar flux of Woods et al. (2009) is defined and absorption and dissociation cross sections are calculated in "p__UV.f90".

(2) The radiative transfer is solved and the optical depth is calculated in "p__photochem_opticaldepth.f90". Ionization and dissociation rates, reaction rate coefficients and production and loss rates of each species are calculated in "p__photochem_rate.f90". The vertical diffusion flux is calculated in "p__photochem_transport.f90". The eddy and binary diffusion coefficients are defined in "p__eddy_diffusion.f90" and "p__molecular_diffusion.f90", respectively. "p__photochem_scheme.f90" calculates the Jacobian matrix and advances the timestep using the implicit method.

(3) At last, "p__io.f90" outputs the calculated simulation results.

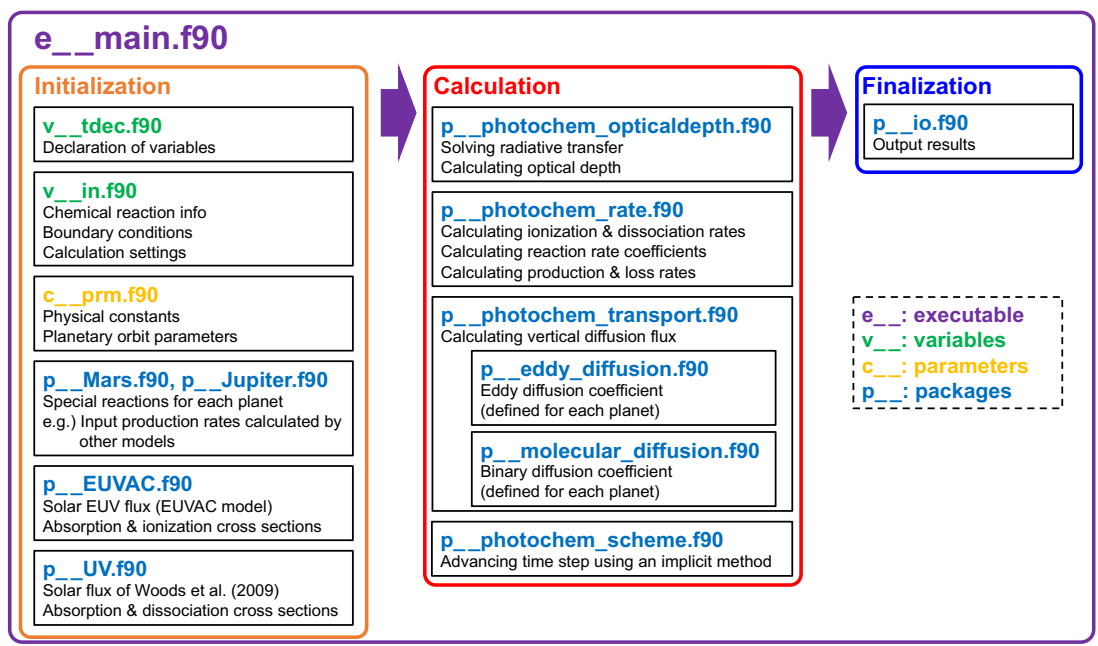

Figure 2 Schematic illustration of the structure of Fortran codes.

**Graphical user interface**

The Python GUI program uses the tkinter package, a standard library of Python to use the Tcl/Tk GUI toolkit (https://docs.python.org/3/library/tkinter.html). The Python GUI allows users to easily and intuitively select a planet, chemical reactions of interest, and run the simulation. An example of GUI is shown in Figure 3, and the operation of GUI is as follows. Once the user runs the Python GUI program, one can select a planet as indicated ("1" in the upper panel of Figure 3). After selecting a planet, one can create a new project directory, or select or rename a project directory that still exists as indicated ("2" in the upper panel of Figure 3). Then the chemical reaction list for the selected planet appears in the window (the lower panel of Figure 3). Chemical reactions and their rate coefficients written in the Python GUI program are automatically converted into Unicode and displayed, making them easy to read. One can select or clear reactions by clicking on the checkbox at each reaction ("3" in the lower panel of Figure 3). By inserting chemical species, reference or label into a search box, only related reactions appear in the window. One can set upper and lower boundary conditions, initial density profiles, vertical grid size, and other calculation settings such as dimension of the simulation, season, latitude, integration time, maximum time step size ("4" in the lower panel of Figure 3). After all the settings are done, one can press "Output f90 module", then the following files are generated in the selected project directory: a Fortran 90 module named "v__in.f90", setting files stored in the directory "settings", and information about production and loss reactions of each chemical species, rate coefficient labels, and Jacobian matrix analyzed by the string parsing function stored in the directory "PLJ_list". The list of chemical species, the number of chemical species, indices, mass, and charge of each chemical species are automatically determined and written in "v__in.f90" at this time. Those text files are read by the Fortran 90 module "v__in.f90". One can also output those files, compile and run the Fortran codes by pressing "Output f90 module & Run model", which requires the installation of an open source software CMake into user's computer ("5" in the lower panel of Figure 3). All the settings and selected reactions are saved, and users can use the same settings and selected reactions the next time they run GUI. At the end of the simulation, users can quickly plot the density profiles by pressing "Plot setting" button ("6" in the lower panel of Figure 3).

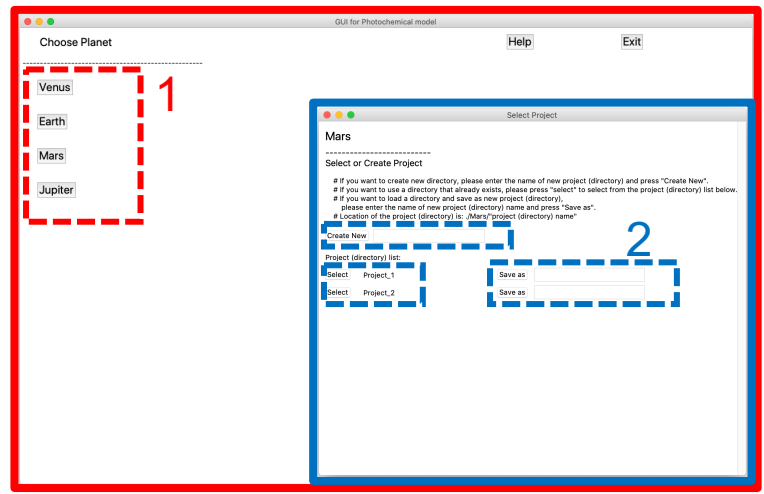

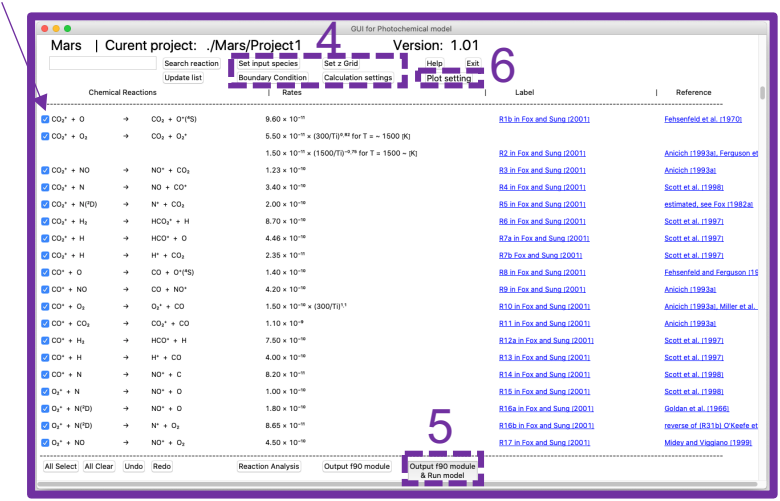

Figure 3 Overview of GUI and instruction of the operation.

**Format of chemical reaction list**

The main feature of PROTEUS is the simple format of chemical reaction list in the Python GUI program and on the GUI. Format of chemical reaction list in the Python GUI program and some examples are illustrated in Figure 4.

```
Examples of chemical reaction list in Python code                                    hv   : Photon
                                                                                     M    : Total atmospheric number density
# Basic format                                                                       e-   : Electron
reaction_rate_list.append(" Reactants  -> Products : Rate coefficient  for T = T1 ~ T2 [K] @ Reference # Label ")    Tn   : Neutral temperature
-----------------------------------------------------------------------------        Ti   : Ion temperature
                                                                                     Te   : Electron temperature
# Photo-ionization and dissociation reactions                                        K0   : Low-pressure-limit rate coefficient
reaction_rate_list.append(" CO2 + hv -> CO2+ + e- : Photoionization ")   } Reaction rates are calculated by using   kinf : High-pressure-limit rate coefficient
reaction_rate_list.append(" CO2 + hv -> CO   + O  : Photodissociation ") }   local photon flux and cross sections

# Normal two-body reactions
reaction_rate_list.append(" O(1D) + N2O -> N2  + O2  : 4.9e-11                      # R75 in Nair et al. [1994] ")
reaction_rate_list.append(" NO    + O3  -> NO2 + O2  : 2.0e-12 * exp(-1400/Tn) # R76 in Nair et al. [1994] ")

# Reaction with several expression of rate coefficient at different temperature ranges
reaction_rate_list.append(" N2+ + O2 -> N2 + O2+ : 5.10e-11 * (300/Ti)^1.16      for T = ~ 1000 [K]       && \
                                                   1.26e-11 * (1000/Ti)^(-0.67)  for T = 1000 ~ 2000 [K]  && \
                                                   2.39e-11                      for T = 2000 ~ [K] @ Scott et al. [1999], Dotan et al. [1997] # R23 in Fox and Sung [2001] ")
# Pressure-dependent three-body reaction
reaction_rate_list.append(" H + O2 + M -> HO2 + M : k0   = 8.8e-32 * (300/Tn)^1.3    && \
                                                    kinf = 7.5e-11 * (300/Tn)^(-0.2)   # Chaffin et al. [2017] ")

# Reaction with unusual rate coefficient equation
reaction_rate_list.append(" N2+ + N2 + M -> N4+ + M : 6.8e-29 * (300/Tn)^2.23 * (1-0.00824*(300/Tn)^0.89) @ Troe [2005] # R31 in Pavlov [2014] ")

# Cluster ion reactions
reaction_rate_list.append(" H+(H2O)4  + H2O + M    -> H+(H2O)5    + M       : 4.6e-28 * (300/Tn)^14   # R41 in Verronen et al. [2016] ")
reaction_rate_list.append(" H+(H2O)4  + CO3-(H2O)2 -> H  + 6H2O + O + CO2  : 6.0e-8 * (300/Tn)^0.5   # R9 in Verronen et al. [2016] ")
```

Figure 4 Format of the chemical reaction list in the Python GUI program.

Any reactions and their rate coefficients are described in the following string format in the Python GUI program. Reaction and rate coefficient are separated by a colon ":", and left- and right-hand side of the reaction are separated by an arrow "->". Chemical species can be written simply as string. For instance, ionized species "$N_2^+$", "$CO_2^+$", and "$H^+(H_2O)_4$" are simply described as "N2+", "CO2+", and "H+(H2O)4", respectively, and electron is described as "e-". Isotope species such as "$^{13}CO_2$" can be written as "^13CO2". Each chemical species and an addition operator "+" or an arrow "->" should be separated by at least one space. PROTEUS also deals with three-body reactions with the expression "M" describing the total atmospheric number density. Temperature-dependent rate coefficient equation can be simply described as string in infix notation. Addition operator "+", subtraction operator "-", multiplication operator "*", division operator "/", exponentiation operator "^" or "**", exponential function "exp()" square root function "sqrt()", neutral, ion and electron temperatures "Tn", "Ti", and "Te", respectively, altitude in km "h", decimal fraction values such as "1.16", integer values such as "300", and values in E notation such as "4.9e-11" can be used in the rate coefficient equation. If there is a temperature range $T_1$-$T_2$ [K] in which the rate coefficient is valid, one can describe the temperature range by "for T = $T_1$ ~ $T_2$ [K]".

Reactions and their rate coefficients selected on GUI are parsed by the string parsing functions in the Python GUI program. Index for each chemical species is automatically determined by the string parsing function, and mass and charge of each chemical species are also automatically identified by the string parsing function. String of each species are automatically divided into constituent elements and mass is calculated by the sum of mass of all the elements, and charge is calculated by counting the number of "+" and "-" in the string of each species. The string parsing function analyzes which reaction produces or lose each chemical species. The rate coefficient expressions written in infix notation are first separated into tokens. Then the order of tokens in infix notation are converted into reverse Polish notation (i.e., postfix notation) and automatically labeled. The Fortran 90 modules calculates the reaction rate coefficient using the labeled tokens arranged in reverse Polish notation. This method allows PROTEUS to process a variety of expression of temperature- and altitude-dependent rate coefficients (as seen in Figure 4) at high computational speed. All the information needed to calculate production rate, loss rate and Jacobian matrix are output as text files, which will be read by Fortran 90 module to apply the information about chemical reactions selected. The number of chemical species and reactions, mass and charge of chemical species, rate coefficient of each reaction and contribution of each reaction to production/loss of each species are automatically applied to Fortran 90 modules by reading those text files. Those features make PROTEUS a flexible photochemical model that can be applied to many planetary atmospheres with different set of chemical reactions.

**Application to planetary atmospheres**
**Mars**

For the application to the Martian atmosphere, parameters of Mars and its orbit are implemented into PROTEUS; The mean distance between Mars and the sun is $r_m$=1.524 AU, the eccentricity is $e$= 0.0934, the solar longitude at perihelion is $L_{s,P}$=250°, tilt angle of the rotational axis is $\varepsilon$=25.2°, the rotational period is $T_p$=88775 sec, the mass of Mars is 6.417×10$^{23}$ kg, and the mean radius of Mars is 3389.5 km (Patel et al., 2002; Williams, 2021).

The cross sections implemented into PROTEUS for the application to Mars are as follows. Ionization cross sections of $CO_2$, $CO$, $O_2$, $N_2$, and $O$ are taken from Schunk and Nagy (2009). Absorption cross sections and quantum yields for calculating dissociation rates of atmospheric molecules are listed in Table A.1. In order to validate PROTEUS, we compared with one-dimensional Martian photochemical model by Chaffin et al. (2017) (hereafter called as C17 model), using the same chemical reactions and their rate coefficients, boundary conditions, temperature and water vapor profiles, and binary and eddy diffusion coefficient profiles. The neutral density profiles simulated by PROTEUS and C17 model are shown in Figure 5. PROTEUS and C17 model are in good agreement except for small differences for $O_3$, $OH$, $HO_2$ and $H_2O_2$. Those differences could be due to the difference in the photo-absorption and dissociation cross sections used in the two models.

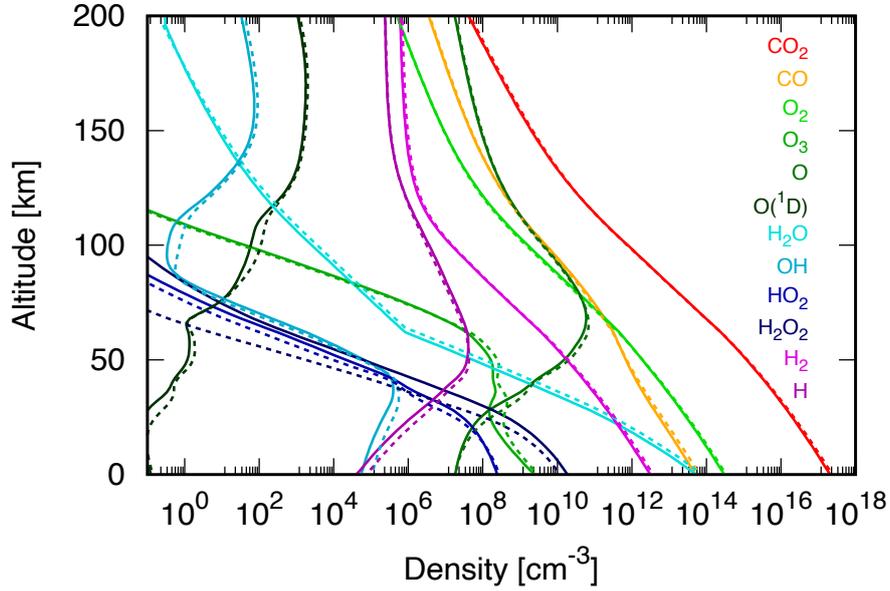

Figure 5 Vertical profiles of neutral density simulated by PROTEUS (solid) and one-dimensional Martian photochemical model by Chaffin et al. (2017) (dashed). The same boundary conditions, chemical reactions and their rate coefficient, binary and eddy diffusion coefficients, and temperature profile, were used in both simulations for validation.

**Jupiter**

For the application to the Jovian atmosphere, parameters of Jupiter and its orbit are implemented into PROTEUS; The mean distance between Jupiter and the sun is $r_m$=5.2 AU, the eccentricity $e$, the solar longitude at perihelion $L_{s,P}$, and tilt angle of the rotational axis $\varepsilon$ are set to zero for simplicity, the rotational period is assumed to be the System III period related to the period of radio burst $T_p$= 35729.71 sec, the mass of Jupiter is 1.898×10$^{27}$ kg, and the equatorial radius of Jupiter is 71492 km (Williams, 2021; Russell et al., 2001).

Ionization cross sections of hydrogen molecule and atom, helium atom, hydrocarbon molecules (CH$_4$, C$_2$H$_2$, C$_2$H$_4$, and C$_2$H$_6$) and metallic atoms (Fe, Mg, Si, and Na) implemented into PROTEUS for the application to the Jovian ionosphere are found in Appendix of Nakamura et al. (2022) and references therein.

PROTEUS has recently been applied to the Jovian ionosphere by Nakamura et al. (2022). Chemical reactions regarding hydrocarbon ion chemistry used in the simulation are described in Nakamura et al. (2022). Ion density profiles calculated by PROTEUS with 218 reactions are shown in Figure 6. Simulated ion density profiles are in good agreement with Kim and Fox (1994), as discussed in Nakamura et al. (2022). Slight differences seen in the shape of profiles of hydrocarbon ions could result from the difference in the initial density profiles of hydrocarbon molecules, which are not indicated in Kim and Fox (1994).

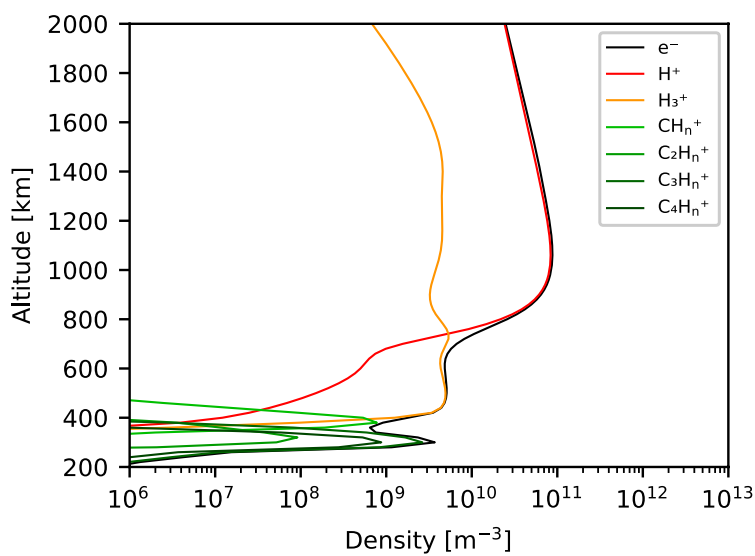

Figure 6 Ion density profiles of the Jovian ionosphere simulated by PROTEUS. Profiles are the same as Figure 3(a) in Nakamura et al. (2022) that used PROTEUS.

**Summary**

We have newly developed a flexible one-dimensional photochemical model named PROTEUS, which consists of a Python GUI program and of Fortran 90 modules. Chemical reactions can be easily implemented into Python code as a simple string format, and users can intuitively select a planet and chemical reactions to be considered in their calculation on GUI. Chemical reactions selected on GUI are automatically analyzed by a string parsing code written in Python, which will be applied to Fortran 90 modules to simulate with selected chemical reactions on a selected planet. This paper presents examples of PROTEUS application to the Martian atmosphere and the Jovian ionosphere, which are in good agreement with previous numerical models. PROTEUS can significantly save time for those who need to develop a new photochemical model; they just need to add chemical reactions in the Python code and just select them on GUI to run a new photochemical model. PROTEUS can be easily extended to other planets and satellites, e.g., Venus, Earth, Titan, and exoplanets in the future.

# Appendix

Table A.1 List of cross sections and quantum yields implemented into PROTEUS.

|  | Species or reactions | Wavelength range | References |
|---|---|---|---|
| $\sigma^a$ | $CO_2$ (absorption) | 0.1254-138.8869 nm | Huestis and Berkowitz (2011)[a] |
|  |  | 138.8913 - 212.7660 nm | Schmidt et al. (2013) |
| $\phi$ | $CO_2 + h\nu \rightarrow CO + O$ | 138.8913 - 212.7660 nm | (Assumed to be 1.0) |
| $\sigma^d$ | $CO_2 + h\nu \rightarrow CO + O(^1D)$ | 0.1 - 138 nm | Huebner and Mukherjee (2015)[b] |
| $\sigma^a$ | $^{13}CO_2$ (absorption) | 138.8913 - 212.7660nm | Schmidt et al. (2013) |
| $\sigma^a$ | $O_2$ (absorption) | 0.99 - 43.5 nm | Huffman (1969)[a] |
|  |  | 49.043646 - 103.066357 nm | Holland et al. (1993)[a] |
|  |  | 103.62 - 107.74 nm | Lee (1955)[a] |
|  |  | 108.75 - 114.95 nm | Ogawa and Ogawa (1975)[a] |
|  |  | 115 - 130.02 nm | Lu et al. (2010)[a] |
|  |  | 130.04 - 175.24 nm | Yoshino et al. (2005)[a] |
|  |  | 175.4 - 204 nm | Minschwaner et al. (1992)[a] |
|  |  | 193 - 245 nm | Yoshino et al. (1992)[a] |
| $\phi$ | $O_2 + h\nu \rightarrow O + O$ | 103 - 242 nm | Burkholder et al. (2015) |
| $\phi$ | $O_2 + h\nu \rightarrow O + O(^1D)$ | 103 - 175 nm | Burkholder et al. (2015) |
| $\sigma^a$ | $H_2O$ (absorption) | 6.2 - 59.04 nm | Chan et al. (1993)[a] |
|  |  | 60.01 - 114.58 nm | Gürtler et al. (1977)[a] |
|  |  | 114.80 - 120.35 nm | Mota et al. (2005)[a] |
|  |  | 120.38 - 139.99 nm | Yoshino et al. (1996, 1997)[a] |
|  |  | 140.00 - 196.00 nm | Chung et al. (2001)[a] |
|  |  | 196.031 - 230.413 nm | Ranjan et al. (2020)[a] |
| $\phi$ | $H_2O + h\nu \rightarrow H + OH$ | 105 nm - | Burkholder et al. (2015) |
| $\phi$ | $H_2O + h\nu \rightarrow H_2 + O(^1D)$ | 105 - 145 nm | Burkholder et al. (2015) |
| $\sigma^a$ | $O_3$ (absorption) | 0.06 - 210 nm | Huebner and Mukherjee (2015)[b] |
|  |  | 213.330 - 1100 nm | Gorshelev et al. (2014) |
|  |  |  | Serdyuchenko et al. (2014) |
| $\phi$ | $O_3 + h\nu \rightarrow O_2 + O(^1D)$ | 220 - 340 nm | Matsumi et al. (2002)[a] |
| $\phi$ | $O_3 + h\nu \rightarrow O_2 + O$ | 220 - 340 nm | (Assumed to be $1 - \phi(O_3 \rightarrow O(^1D))$) |
| $\sigma^a$ | $HO_2$ (absorption) | 190 - 260 nm | Burkholder et al. (2015) |
| $\phi$ | $HO_2 + h\nu \rightarrow OH + O$ | 190 - 260 nm | Burkholder et al. (2015) |
| $\sigma^a$ | $H_2O_2$ (absorption) | 121.33 - 189.70 nm | Schürgers and Welge (1968)[a] |
| | | 190.00 - 255.00 nm | Burkholder et al. (2015) |
| $\phi$ | $H_2O_2 + h\nu \rightarrow HO_2 + H$ | 121 - 230 nm | Burkholder et al. (2015) |
| $\phi$ | $H_2O_2 + h\nu \rightarrow OH + OH$ | 121 - 340 nm | Burkholder et al. (2015) |
| $\sigma^a$ | OH (absorption) | 0.06 - 282.3 nm | Huebner and Mukherjee (2015)[b] |
| $\sigma^d$ | $OH + h\nu \rightarrow H + O$ | 124.5 - 261.65 nm | Huebner and Mukherjee (2015)[b] |
| $\sigma^d$ | $OH + h\nu \rightarrow H + O(^1D)$ | 93 - 511.4 nm | Huebner and Mukherjee (2015)[b] |
| $\sigma^a$ | $H_2$ (absorption) | 0.1 - 110.86 nm | Huebner and Mukherjee (2015)[b] |
| $\sigma^d$ | $H_2 + h\nu \rightarrow H + H$ | 84.48 - 110.86 nm | Huebner and Mukherjee (2015)[b] |
| $\sigma^a$ | $N_2$ (absorption) | 0.1 - 103.8 nm | Huebner and Mukherjee (2015)[b] |
| $\sigma^d$ | $N_2 + h\nu \rightarrow N + N$ | 51.96 - 103.8 nm | Huebner and Mukherjee (2015)[b] |
| $\sigma^a$ | NO (absorption) | 0.1 - 191 nm | Huebner and Mukherjee (2015)[b] |
| $\sigma^d$ | $NO + h\nu \rightarrow N + O$ | 0.1 - 191 nm | Huebner and Mukherjee (2015)[b] |
| $\sigma^a$ | $NO_2$ (absorption) | 0.06 - 238 nm | Huebner and Mukherjee (2015)[b] |
| | | 238.08219 - 666.57808 nm | Vandaele et al. (1998)[a] |
| $\sigma^d$ | $NO_2 + h\nu \rightarrow NO + O(^1D)$ | 108 - 243.88 nm | Huebner and Mukherjee (2015)[b] |
| $\phi$ | $NO_2 + h\nu \rightarrow NO + O$ | 108 - 238 nm | Huebner and Mukherjee (2015)[b] |
| | | 239 - 300 nm | (Assumed to be 1) |
| | | 300 - 422 nm | Burkholder et al. (2015) |
| $\sigma^a$ | $NO_3$ (absorption) | 400 - 691 nm | Wayne et al. (1991)[a] |
| $\phi$ | $NO_3 + h\nu \rightarrow NO_2 + O$ | 400 - 640 nm | Johnston et al. (1996)[a] |
| $\phi$ | $NO_3 + h\nu \rightarrow NO + O_2$ | 586 - 640 nm | Johnston et al. (1996)[a] |
| $\sigma^a$ | $N_2O$ (absorption) | 16.8 - 59.0 nm | Hitchcock et al. (1980)[a] |
| | | 60.0 - 99.9 nm | Cook et al. (1968)[a] |
| | | 108.20 - 122.18 nm | Zelikoff et al. (1953)[a] |
| | | 122.25 - 172.88 nm | Rabalais et al. (1971)[a] |
| | | 173 - 210 nm | Selwyn et al. (1977)[a] |
| $\phi$ | $N_2O + h\nu \rightarrow N_2 + O(^1D)$ | 140 - 230 nm | Burkholder et al. (2015) |
| $\sigma^a$ | $N_2O_5$ (absorption) | 152 - 198 nm | Osborne et al. (2000)[a] |
| | | 200 - 260 nm | Burkholder et al. (2015) |
| | | 260 - 410 nm | Burkholder et al. (2015) |
| $\phi$ | $N_2O_5 + h\nu \rightarrow NO_3 + NO_2$ | 248 - 410 nm | Burkholder et al. (2015) |
| $\phi$ | $N_2O_5 + h\nu \rightarrow NO_3 + NO + O$ | 152 - 289 nm | Burkholder et al. (2015) |
| $\sigma^a$ | $HNO_2$ (absorption) | 184 - 396 nm | Burkholder et al. (2015) |
| $\phi$ | $HNO_2 + h\nu \rightarrow NO + OH$ | All | Burkholder et al. (2015) |
| $\sigma^a$ | HNO$_3$ (absorption) | 192 - 350 nm | Burkholder et al. (2015) |
| $\phi$ | HNO$_3$ + $h\nu$ → HNO$_2$ + O | 193 - 260 nm | Estimated[c] |
| $\phi$ | HNO$_3$ + $h\nu$ → HNO$_2$ + O($^1$D) | 193 - 222 nm | Estimated[c] |
| $\phi$ | HNO$_3$ + $h\nu$ → OH + NO$_2$ | 193 - 350 nm | Estimated[c] |
| $\sigma^a$ | HO$_2$NO$_2$ (absorption) | 190 - 280 nm | Burkholder et al. (2015) |
| | | 280 - 350 nm | Burkholder et al. (2015) |
| $\phi$ | HO$_2$NO$_2$ + $h\nu$ → HO$_2$ + NO$_2$ | 190 - 350 nm | Burkholder et al. (2015) |
| $\phi$ | HO$_2$NO$_2$ + $h\nu$ → OH + NO$_3$ | 190 - 350 nm | Burkholder et al. (2015) |
| $\sigma^a$ | H$_2$CO (absorption) | 224.56 - 376 nm | Meller and Moortgat (2000)[a] |
| $\phi$ | H$_2$CO + $h\nu$ → H$_2$ + CO | 250 - 360 nm | Burkholder et al. (2015) |
| $\phi$ | H$_2$CO + $h\nu$ → H + HCO | 250 - 360 nm | Burkholder et al. (2015) |

$\sigma^a$: Absorption cross section, $\sigma^d$: dissociation cross section, $\phi$: quantum yield, [a]: data files are taken from The MPI-Mainz UV/VIS Spectral Atlas (Keller-Rudek et al., 2013), [b]: data files are taken from PHIDRATES (Huebner and Mukherjee, 2015), [c]: quantum yields for each photolysis reaction of HNO$_3$ were estimated by quantum yield of each product (OH, O, and O($^1$D)) obtained by Johnston et al. (1974), Turnipseed et al. (1992), and Margitan and Watson (1982).

**Acknowledgements**

**Reference**

Arney, G., Domagal-Goldman, S. D., Meadows, V. S., Wolf, E. T., Schwieterman, E., Charnay, B., Claire, M., Hébrard, E., and Trainer, M. G. (2016). The Pale Orange Dot: The Spectrum and Habitability of Hazy Archean Earth, *Astrobiology*, 16:11, 873-899 doi:10.1089/ast.2015.1422.

Burkholder, J. B., Sander, S. P., Abbatt, J., Barker, J. R., Huie, R. E., Kolb, C. E., Kurylo, M. J., Orkin, V. L., Wilmouth, D. M., and Wine, P. H. (2015). Chemical Kinetics and Photochemical Data for Use in Atmospheric Studies, Evaluation No. 18, *JPL Publication 15-10*, Jet Propulsion Laboratory, Pasadena, http://jpldataeval.jpl.nasa.gov.

Chaffin, M. S., Deighan, J., Schneider, N. M., and Stewart, A. I. F. (2017). Elevated atmospheric escape of atomic hydrogen from Mars induced by high-altitude water, *Nature Geoscience*, 10, 174–178, doi:10.1038/ngeo2887.

Chan, W. F., Cooper, G., and Brion, C. E. (1993). The electronic spectrum of water in the discrete and continuum regions. Absolute optical oscillator strengths for photoabsorption (6-200 eV), *Chem. Phys.*, 178, 387-401, doi:10.1016/0301-0104(93)85078-M.

Chamberlin, P. C., Eparvier, F. G., Knoer, V., Leise, H., Pankratz, A., Snow, M., et al. (2020). The flare irradiance spectral model-version 2 (FISM2), *Space Weather*, 18, e2020SW002588, doi:10.1029/2020SW002588

Chung, C.-Y., Chew, E. P., Cheng, B.-M., Bahou, M., and Lee, Y.-P. (2001). Temperature dependence of absorption cross-section of $H_2O$, HDO, and $D_2O$ in the spectral region 140-193 nm, *Nuclear Instruments and Methods in Physics Research Section A: Accelerators, Spectrometers, Detectors and Associated Equipment*, 467–468, Part 2, 1572-1576, doi.org/10.1016/S0168-9002(01)00762-8.

Cook, G. R., Metzger, P. H., and Ogawa, M. (1968). Photoionization and absorption coefficients of $N_2O$, *J. Opt. Soc. Am.*, 58, 129-136, doi:10.1364/JOSA.58.000129.

Damian, V., Sandu, A., Damian, M., Potra, F., and Carmichael, G. R. (2002). The kinetic preprocessor KPP - a software environment for solving chemical kinetics, *Computers & Chemical Engineering*, 26, 11, 1567-1579, doi:10.1016/S0098-1354(02)00128-X.

Dotan, I., Hierl, P. M., Morris, R. A., and Viggiano, A. A. (1997). Rate constants for the reactions of N+ and N2+ with O2 as a function of temperature (300 - 1800 K), *Int. J. Mass Spectrom. Ion Proc.*, 167/168, 223-230, doi:10.1016/S0168-1176(97)00077-3.

Fox, J. L., and Sung, K. Y. (2001). Solar activity variations of the Venus thermosphere/ionosphere, *J. Geophys. Res.*, 106(A10), 21305-21335, doi:10.1029/2001JA000069.

Gorshelev, V., Serdyuchenko, A., Weber, M., Chehade, W., and Burrows, J. P., (2014). High spectral resolution ozone absorption cross-sections – Part 1: Measurements, data analysis and comparison with previous measurements around 293 K, *Atmos. Meas. Tech.*, 7, 609–624,

2014, doi.org/10.5194/amt-7-609-2014.

Gürtler, P., Saile, V., and Koch, E. E. (1977). Rydberg series in the absorption spectra of $H_2O$ and $D_2O$ in the vacuum ultraviolet, *Chem. Phys. Lett.*, 51, 386-391, doi:10.1016/0009-2614(77)80427-2.

Hitchcock, A. P., Brion, C. E., and van der Wiel, M. J. (1980). Absolute oscillator-strengths for valence-shell ionic photofragmentation of $N_2O$ and $CO_2$ (8-75 eV), *Chem. Phys.*, 45, 461-478, doi:10.1016/0301-0104(80)87015-7.

Holland, D. M. P., Shaw, D. A., McSweeney, S. M., MacDonald, M. A., Hopkirk, A., Hayes, M. A. (1993). A study of the absolute photoabsorption, photoionization and photodissociation cross sections and the photoionization quantum efficiency of oxygen from the ionization threshold to 490 Å, *Chemical Physics*, 173, 2, 315-331, doi:10.1016/0301-0104(93)80148-3.

Huebner, W.F., and Mukherjee, J. (2015). Photoionization and photodissociation rates in solar and blackbody radiation fields, *Planetary and Space Science*, 106, 11-45, doi:10.1016/j.pss.2014.11.022.

Huestis, D. L., and Berkowitz, J. (2011). Critical evaluation of the photoabsorption cross section of $CO_2$ from 0.125 to 201.6 nm at room temperature, *Advances in Geosciences*, 229-242, doi:10.1142/9789814355377_0018.

Huffman, R. E. (1969). Absorption cross-sections of atmospheric gases for use in aeronomy, *Canadian Journal of Chemistry*, 47 (10), 1823-1834 doi:10.1139/v69-298

Johnston, H. S., Chang, S.-G., and Whitten, G. (1974). Photolysis of nitric acid vapor, *J. Phys. Chem.*, 78, 1-7, doi:10.1021/j100594a001.

Johnston, H. S., Davis, H. F., and Lee, Y. T. (1996). $NO_3$ photolysis product channels: Quantum yields from observed energy thresholds, *J. Phys. Chem.*, 100, 4713-4723, doi:10.1021/jp952692x.

Kasting, J. F., Liu, S. C., and Donahue, T. M. (1979). Oxygen levels in the prebiological atmosphere, *J. Geophys. Res.*, 84(C6), 3097–3107, doi:10.1029/JC084iC06p03097.

Keller-Rudek, H., Moortgat, G. K., Sander, R., and Sörensen, R. (2013). The MPI-Mainz UV/VIS spectral atlas of gaseous molecules of atmospheric interest, *Earth Syst. Sci. Data*, 5, 365–373, doi:10.5194/essd-5-365-2013.

Kim, Y. H., and Fox, J. L. (1994). The chemistry of hydrocarbon ions in the Jovian ionosphere, *Icarus*, 112, 310–325, doi:10.1006/ icar.1994.1186.

Krasnopolsky, V. A., (2009). A photochemical model of Titan's atmosphere and ionosphere, *Icarus*, 201, 1, 226-256, doi:10.1016/j.icarus.2008.12.038.

Krasnopolsky, V. A. (2012). A photochemical model for the Venus atmosphere at 47–112km, Icarus, 218, 1,230-246, doi:10.1016/j.icarus.2011.11.012.

Lee, P. (1955). Photodissociation and photoionization of oxygen ($O_2$) as inferred from measured absorption coefficients, *J. Opt. Soc. Am.*, 45, 703-709, doi:10.1364/JOSA.45.000703.

Liou, K. N. (2002). *An introduction to atmospheric radiation (Vol. 84)*, Elsevier.

Lu, H.-C., Chen, K.-K., Chen, H.-F., Cheng, B.-M., and Ogilvie, J. F. (2010). Absorption cross section of molecular oxygen in the transition $E^3\Sigma_u^-$, v = 0 - $X^3\Sigma_g^-$, v = 0 at 38 K, *Astronom. Astrophys.*, 520, A19, 1-4, doi:10.1051/0004-6361/201013998.

Margitan, J. J., and Watson, R. T. (1982). Kinetics of the reaction of hydroxyl radicals with nitric acid, *J. Phys. Chem.*, 1982, 86, 3819-3824, doi:10.1021/j100216a022.

Matsumi, Y., Comes, F. J., Hancock, G., Hofzumahaus, A., Hynes, A. J., Kawasaki, M., and Ravishankara, A. R. (2002). Quantum yields for production of O($^1$D) in the ultraviolet photolysis of ozone: Recommendation based on evaluation of laboratory data, *J. Geophys. Res.*, 107(D3), ACH1-1 - ACH1-12, doi:10.1029/2001JD000510.

Meller, R., and Moortgat, G. K. (2000). Temperature dependence of the absorption cross sections of formaldehyde between 223 and 323 K in the wavelength range 225-375 nm, *J. Geophys. Res.*, 105(D6), 7089-7101, doi:10.1029/1999JD901074.

Minschwaner, K., Anderson, G. P., Hall, L. A., and Yoshino, K. (1992). Polynomial coefficients
20

for calculating O2 Schumann-Runge cross sections at 0.5 cm$^{-1}$ resolution, *J. Geophys. Res.*, 97(D9), 10103–10108, doi:10.1029/92JD00661.

Mota, R., Parafita, R., Giuliani, A., Hubin-Franskin, M.-J., Lourenço, J. M. C., Garcia, G., Hoffmann, S. V., Mason, M. J., Ribeiro, P. A., Raposo, M., and Limão-Vieira, P. (2005). Water VUV electronic state spectroscopy by synchrotron radiation, *Chem. Phys. Lett.*, 416, 152-159, doi:10.1016/j.cplett.2005.09.073.

Nair, H., Allen, M., Anbar, A. D., Yung, Y. L., and Clancy, R. T., (1994). A Photochemical Model of the Martian Atmosphere, *Icarus*, 111, 1, 124-150, doi:/10.1006/icar.1994.1137.

Nakamura, Y., Terada, K., Tao, C., Terada, N., Kasaba, Y., Leblanc, F., et al. (2022). Effect of meteoric ions on ionospheric conductance at Jupiter. *Journal of Geophysical Research: Space Physics*, 127, e2022JA030312, doi:10.1029/2022JA030312.

Ogawa, S., and Ogawa, M. (1975). Absorption cross sections of $O_2(a^1\Delta_g)$ and $O_2(X^3\Sigma_g^-)$ in the region from 1087 to 1700 Å", *Can. J. Phys.*, 53, 1845-1852, doi:10.1139/p75-236.

Osborne, B. A., Marston, G., Kaminski, L., Jones, N. C., Gingell, J. M., Mason, N., Walker, I. C., Delwiche, J., and Hubin-Franskin, M.-J. (2000). Vacuum ultraviolet spectrum of dinitrogen pentoxide, *J. Quant. Spectrosc. Radiat. Transfer*, 64, 67-74doi:10.1016/S0022-4073(99)00104-1.

Patel, M. R., Zarnecki, J. C., and Catling, D. C. (2002). Ultraviolet radiation on the surface of Mars and the Beagle 2 UV sensor, *Planetary and Space Science*, 50, 9, 915-927, doi:10.1016/S0032-0633(02)00067-3.

Pavlov, A. V. (2014). Photochemistry of Ions at D-region Altitudes of the Ionosphere: A Review, *Surv. Geophys.*, 35:259–334, doi:10.1007/s10712-013-9253-z.

Rabalais, J. W., McDonald, J. M., Scherr, V., and McGlynn, S. P. (1971). Electronic spectroscopy of isoelectronic molecules. II. Linear triatomic groupings containing sixteen valence electrons, *Chem. Rev.*, 71, 73-108, doi:10.1021/cr60269a004.

Ranjan, S., Schwieterman, E. W., Harman, C., Fateev, A., Sousa-Silva, C., Seager, S., and Hu, R. (2020). Photochemistry of anoxic abiotic habitable planet atmospheres: Impact of new $H_2O$

cross-sections, *Astrophys. J.*, 896, 148, doi:10.3847/1538-4357/ab9363.

Richards, P. G., Fennelly, J. A., and Torr, D. G. (1994). EUVAC: A solar EUV flux model for aeronomic calculations, *J. Geophys. Res.*, 99(A5), 8981-8992, doi:10.1029/94JA00518.

Richards, P. G., Woods, T. N., and Peterson, W. K. (2006). HEUVAC: A new high resolution solar EUV proxy model, *Advances in Space Research*, 37, 2, 315-322, doi.org/10.1016/j.asr.2005.06.031.

Schunk, R. W., and Nagy, A. F. (2009). *Ionospheres, 2nd edition*, Cambridge University Press.

Schürgers, M., and Welge, K. H. (1968). Absorptions koeffizient von $H_2O_2$ und $N_2H_4$ zwischen 1200 und 2000 Å, *Z. Naturforsch.*, 23a, 1508-1510, doi:10.1515/zna-1968-1011.

Scott, G. B. I., Fairley, D. A., Milligan, D. B., Freeman, C. G., and McEwan, M. J. (1999). Gas phase reactions of some positive ions with atomic and molecular oxygen and nitric oxide at 300 K, *J. Phys. Chem. A*, 103, 7470-7473, doi:10.1021/jp9913719.

Selwyn, G., Podolske, J., and Johnston, H. S. (1977). Nitrous oxide ultraviolet absorption spectrum at stratospheric temperatures, *Geophys. Res. Lett.*, 4, 427-430, doi:10.1029/GL004i010p00427.

Serdyuchenko, A., Gorshelev, V., Weber, M., Chehade, W., and Burrows, J. P. (2014). High spectral resolution ozone absorption cross-sections – Part 2: Temperature dependence, *Atmos. Meas. Tech.*, 7, 625–636, doi:10.5194/amt-7-625-2014.

Troe, J. (2005). Temperature and pressure dependence of ion–molecule association and dissociation reactions: the $N_2^+ + N_2$ (+ M) $\Leftrightarrow N_4^+$ (+ M) reaction, *Phys. Chem. Chem. Phys.*, 7, 1560-1567, doi:10.1039/B417945P.

Turnipseed, A. A., Vaghjiani, G. L., Thompson, J. E., and Ravishankara, A. R. (1992). Photodissociation of $HNO_3$ at 193, 222, and 248 nm: Products and quantum yields, *J. Chem. Phys.*, 1992, 96, 5887-5895, doi:10.1063/1.462685.

Vandaele, A. C., Hermans, C., Simon, P. C., Carleer, M., Colins, R., Fally, S., Mérienne, M. F., Jenouvrier, A., and Coquart, B. (1998). Measurements of the $NO_2$ absorption cross-sections from 42000 cm-1 to 10000 cm-1 (238-1000 nm) at 220 K and 294 K, *J. Quant. Spectrosc.*

*Radiat. Transfer*, 59, 171-184, doi:10.1016/S0022-4073(97)00168-4.

Verronen, P. T., Andersson, M. E., Marsh, D. R., Kovács, T., and Plane, J. M. C. (2016), WACCM-D—Whole Atmosphere Community Climate Model with D-region ion chemistry, *J. Adv. Model. Earth Syst.*, 8, 954–975, doi:10.1002/2015MS000592.

Wayne, R. P., Barnes, I., Burrows, J. P., Canosa-Mas, C. E., Hjorth, J., Le Bras, G., Moortgat, G.K., Perner, D., Poulet, G., Restelli, G., and Sidebottom, H. (1991). The nitrate radical: Physics, chemistry, and the atmosphere, *Atmos. Environ.*, 25A, 1-203, doi:10.1016/0960-1686(91)90192-A.

Williams, D. (2021). Mars Fact Sheet, *NASA Goddard Space Flight Center*, retrieved on 16 October 2022.

Woods, T. N., Chamberlin, P. C., Harder, J. W., Hock, R. A., Snow, M., Eparvier, F. G., Fontenla, J., McClintock, W. E., and Richard, E. C. (2009). Solar Irradiance Reference Spectra (SIRS) for the 2008 Whole Heliosphere Interval (WHI), *Geophys. Res. Lett.*, 36, L01101, doi:10.1029/2008GL036373.

Yoshida, T., Aoki, S., Ueno, Y., Terada, N., Nakamura, Y., Shiobara, K., Yoshida, N., Nakagawa, H., Sakai, S., and Koyama, S. (2022). Strong depletion of $^{13}$C in CO induced by photolysis of $CO_2$ in the Martian atmosphere calculated by a photochemical model, *Planetary Science Journal*, submitted.

Yoshino, K., Esmond, J. R., Cheung, A. S.-C, Freeman, D. E., and Parkinson, W. H. (1992). High resolution absorption cross sections in the transmission window region of the Schumann-Runge bands and Herzberg continuum of $O_2$, *Planetary and Space Science*, 40, 2–3, 185-192, doi:10.1016/0032-0633(92)90056-T.

Yoshino, K., Esmond, J. R., Parkinson, W. H., Ito, K., and Matsui, T. (1996). Absorption cross section measurements of water vapor in the wavelength region 120 to 188 nm, *Chem. Phys.*, 211, 387-391, doi:10.1016/0301-0104(96)00210-8.

Yoshino, K., Esmond, J. R., Parkinson, W. H., Ito, K., Matsui, T. (1997). Absorption cross section measurements of water vapor in the wavelength region 120 nm to 188 nm (Chem. Phys. 211 (1996) 387–391). *Chemical Physics*, 215. 429-430, doi:10.1016/s0301-0104(96)00381-3.

Yoshino, K., Parkinson, W. H., Ito, K., and Matsui, T. (2005). Absolute absorption cross-section measurements of Schumann-Runge continuum of $O_2$ at 90 and 295 K, *J. Mol. Spectrosc.*, 229, 238-243, doi:10.1016/j.jms.2004.08.020.

Zahnle, K., Claire, M. and Catling, D. (2006). The loss of mass-independent fractionation in sulfur due to a Palaeoproterozoic collapse of atmospheric methane, *Geobiology*, 4: 271-283, doi:10.1111/j.1472-4669.2006.00085.x.

Zelikoff, M., Watanabe, K., and Inn, E. C. Y. (1953). Absorption coefficients of gases in the vacuum ultraviolet. Part II. Nitrous oxide, *J. Chem. Phys*., 21, 1643-1647, doi:10.1063/1.1698636.